\documentstyle[prl,aps,floats,epsf]{revtex}
\newcommand{\BT}{\mbox{BaTiO$_{3}$}}
\newcommand{\tg}{\mbox{$t_{2g}$}}
\newcommand{\tb}{\mbox{\tiny $\bullet$}}
\begin{document}
\preprint{version 1}
\draft
\title{Formation of localized hole states in complex oxides}
\author{H.~Donnerberg, S.~T{\"o}bben, A.~Birkholz}

\address{University of Osnabr\"uck, FB Physik, D-49069 Osnabr\"uck, FRG}
\date{\today}
\maketitle
\begin{abstract}
Defect electrons (holes) play an important role in most technologically important
complex oxides.
In this contribution we present the first detailed characterization of localized hole states
in such materials. Our investigations employ advanced
embedded-cluster calculations which consistently include electron correlations
and defect-induced lattice relaxations. This is necessary in order to account
for the variety of possible hole-state manifestations.
\end{abstract}

\section{Introduction}
\label{int}
The basic structural building units of solid oxides are
$\rm MO_6$ metal-oxygen octahedra and, possibly,
additional $\rm MO_4$ tetrahedra.
The complete crystal structure
is built up of corner-, face- and/or edge-sharing connections of these structural
elements. Additional cations (A) can be incorporated at interstitial lattice
sites. These are increasingly formed, the more open-structured the whole
network of $\rm MO_n$ units appears to be. Generally, open crystal structures
imply high formal M-cation charge states (referring to formal $\rm O^{2-}$
anions) leading to mixed ionic-covalent (or semi-ionic)
material properties. The M-cations are in most cases transition-metal (TM) ions.
Examples of such complex
materials are given by $\rm AMO_3$ perovskite-structured oxides such as
barium titanate (\BT). Perovskite oxides are frequently ferroelectric
and possess important electrooptic applications based on the photorefractive
effect. Also high-T$_C$ oxides resemble the perovskite structure.

Charge carriers, either valence-band (VB) holes or con\-duc\-tion-band (CB)
electrons, are created by doping with impurities,
annealing treatments or by light-induced charge-transfer excitations
which, for example, take place during photorefractive  processes
in the appropriate oxides. Combined optical-absorption- and
electron-spin-resonance measurements (Photo-ESR) \cite{PJK92}
proved that the created and afterwards trapped
holes influence the light-induced charge-transfer reactions in photorefractive
{\BT}. These holes, which are either para- or diamagnetic, probably induce the
observed sublinear dependence of photoconductivity on the light intensity (e.g.
\cite{Ho89}). A further example highlighting the relevance of holes refers to
high-T$_C$ oxides: Pairs of doped holes give rise to
superconductivity in these materials. The possible pairing scenarios
are still a matter of active debate.

The present work initiates a systematic and detailed
characterization of localized hole states in complex oxides.
Our investigations employ real-space embedded-cluster calculations, which
consistently combine electron correlations and defect-induced
lattice relaxations. This procedure is indispensable in order to
predict the richness of possible hole states. Here, we demonstrate simulations
for trapped holes in {\BT}, but many results, ranging from stabilization
of cationic charge states to the formation of bipolarons, can be extrapolated to
other oxides.

The trapping of holes at acceptor defects leads to a localization of hole
states, which is further aided by defect-induced lattice distortions.
In near-insulating high-T$_C$ oxides the antiferromagnetic order supports
hole localizations. Principally, there is
'on-acceptor' and 'off-acceptor' hole trapping. In the first case the
trapped hole localizes at the acceptor site (thereby increasing the acceptor
charge state by one positive unit, i.e. $\rm M^{+n} + h^{\tb}\longrightarrow
M^{+(n+1)}$), and in the second case the hole remains at the oxygen ligands.
Further, the off-acceptor case allows to classify hole states according to
their degree of delocalization, i.e. complete localization at exactly one
ligand anion, intermediate localization at two neighboring oxygen ions
($\rm V_k$ centers), and delocalization over more than two oxygen ligands.

Which hole-localization type is favored depends on the ionicity of the
host system, but also on the incorporation site and
electronic structure of the
acceptor. For example, at Ba-site acceptors localized
off-acceptor holes are likely to be preferred over delocalized
species due to large interionic separations \cite{VSK96}.
Further, the simultaneous binding of two holes at one acceptor
facilitates the formation of small intersite bipolarons representing
negative-$U$ centers. Such species could explain the diamagnetic hole centers in
photorefractive oxides \cite{PJK92}; intersite bipolarons have also been
considered as possible superconductivity-carrying bosons in high-T$_C$ oxides
(see \cite{AM94}, for details).
Earlier work on bipolarons in oxide materials dealt exclusively with electron type species.
We quote in this context the pioneering work of P.~W.~Anderson \cite{An75} on the 
formation of negative-$U$ centers in amorphous solids. Schlenker \cite{Sc85} investigated 
small electron type bipolarons in titanium oxides.

\section{Methods}
\label{meth}
Our investigations employ real-space embedded-cluster calculations (ECC), in which the 
crystal is devided into a central and quantum chemically treated defect cluster and into
the embedding lattice which is modeled more approximately.
The ECC discussed in this publication deal with
the trapping of holes at Ti-site acceptors in \BT. Our calculations are
based on 21-atom clusters $\rm MO_6Ba_8Ti_6$ centered around
an acceptor-oxygen octahedron MO$_6$, see also \cite{DBa96}.
All simulations assume the perfect-crystal structure of cubic 
{\BT}. This simplification may be justified due to the observation that
all possible ferroelectric distortions of the material are small compared with usual 
defect-induced lattice relaxations.

The molecular-orbital ansatz for the central MO$_6$ complex employs
Gaussian type basis functions with split-valence quality for the acceptor
metal impurity and its
oxygen ligands; the oxygen basis set is further augmented by polarizing $d$-functions.
Bare effective core potentials \cite{HW85} are used to model the localizing
ion-size effects of the outer Ba- and Ti-cations. The {\it ab initio} level
of our calculations covers Hartree-Fock (HF) theory,
M{\o}ller-Plesset perturbation theory to second order (MP2)
and density functional theory (DFT). All quantum cluster simulations have been performed on the 
basis of the quantum-chemical CADPAC code \cite{CAD}.
Within DFT two choices are used to approximate the
exchange-correlation functional, i.e. the local spin density ansatz of Vosko,
Wilk and Nusair \cite{VWN80} (VWN-LSDA) and the generalized-gradient-approximation 
functional involving the Becke exchange term \cite{Be88} and the correlation
functional derived by Lee, Yang and Parr \cite{LYP88}. This functional, abbreviated as 'BLYP',
improves on LDA.  

\begin{figure}[htb]
\begin{center}
\begin{minipage}{10cm}
\epsfxsize10cm
\epsffile{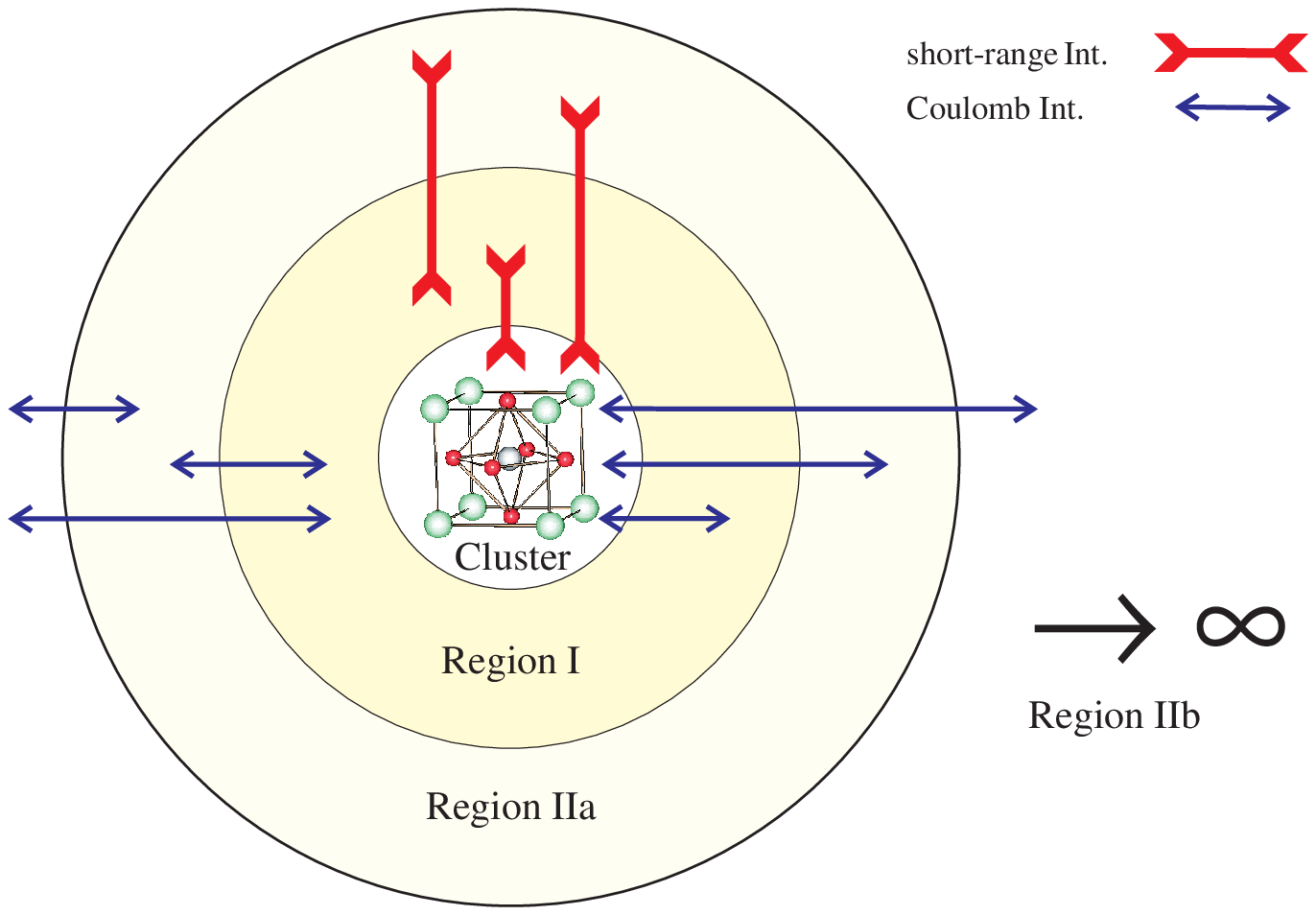}
\end{minipage}
\vspace{0.3cm}
\caption{\label{ecr} \baselineskip12pt
Visualization of embedded-cluster calculations. The quantum cluster is embedded in a classically
treated lattice consisting of the regions I, IIa and IIb. The arrows indicate the various existing
interactions between the crystal regions.}
\end{center}
\end{figure}

The modeling of the embedding lattice and of cluster-lattice interactions
employs a shell-model pair-potential description (see also Refs.
\cite{DBa96,DB95,Do95}). The shell-model parameters have been taken from the earlier work
of Lewis and Catlow \cite{LC86}. It is noted that the shell model  takes
electronic polarization contributions of the embedding lattice ions reasonably well into
account. This feature is particularly important with respect to the highly polarizable
oxygen anions.  Figure \ref{ecr} displays the embedding scheme. 
As in any classical 
Mott-Littleton defect calculation the (embedding) crystal lattice is divided into three regions:
region I which is explicitly equilibrated according to the underlying pair potentials, the 
interface region IIa and region IIb.  Region II(a and b) is treated as a polarizable continuum 
using the harmonic approximation for the region-II self-energy. 
Whereas the interactions between the cluster and region I, on the one hand, and region IIa,
on the other hand, are included explicitly, 
all region-IIb species feel only the effective defect 
charge of the central defect cluster.  Details of 
this well-known description are given in Ref. \cite{CM82}. All lattice calculations 
employ the CASCADE code \cite{Le83}. 

The total energy of the complete model crystal is given by the expression:
\begin{eqnarray}
\label{eecr}
\rm E(\rho,R_c,R_r)= \min_{\rho,R_c,R_r} \{E^{QM}_{Clust}(\rho,R_c)+
\rm E^{SM}_{Env}(R_r)+E^{Int}(\rho,R_c,R_r)\}\,.
\end{eqnarray}
$\rm \rho,~R_c ~and ~R_r$ denote the cluster electron density, the coordinates of the cluster
nuclei and the positions of the shell-model species in the embedding environment, respectively.
$\rm E^{QM}_{Clust}$ represents the quantum mechanically calculated cluster self-energy, 
$\rm E^{SM}_{Env}$ the Mott-Littleton type shell-model energy of the embedding environment and 
$\rm E^{Int}$ is the cluster-lattice interaction energy. The required minimization in 
Eq.(\ref{eecr}) is performed using the programs CADPAC (variation of $\rho$ with fixed $\rm R_r$)
and CASCADE (variation of $\rm R_r$ with fixed $\rm R_c$ and $\rho$). In order to do cluster 
geometry optimizations which are consistent with a predetermined embedding crystal 
lattice (represented by a 
point-charge field) an additional program has been written which updates the total cluster
energies and gradients, as calculated by any quantum chemical program such as CADPAC,
by adding the appropriate short-range pair-potential contributions due to the  
interactions between cluster ions and embedding lattice species. With these updates the
program carries through the cluster geometry optimization using a variable-metric
(quasi-Newton) minimization algorithm. The total minimization procedure is completed, when the
main cycle consisting of the alternating lattice- and cluster-equlibration subcycles converges.
It is finally noted, that we have neglected the full 
cluster multipole consistency during the embedding-lattice relaxation step (see \cite{VHH84}, 
for details), since, in our simulations, we did not observe any appreciable relaxation and
energy changes due to cluster multipole contributions deviating from formal-charge models.

The present approach proves to be the only successful one which can 
consistently integrate
the local electronic structure and large-scale lattice-distortion effects.
In particular, we find the combination of DFT and of our embedding scheme to be sufficiently
accurate and flexible in order to realistically account for the variety
of possible hole states. We stress that cutting off the electronic structure beyond the
cluster boundary does not principally invalidate our results. This is due to the fact that 
embedded-cluster calculations formally refer to a localized-orbital point of view
(e.g. see \cite{Ku78,DBa96} for related discussions) and not to the extended one-electron 
eigenstates of the crystal. Moreover, all hole states considered in
this contribution may be assumed to be localized within the cluster region due
to the presence of the acceptor defect and due to pronounced lattice deformations 
(small-polaron effect).
Further indications towards a reliability of our present results are provided by our earlier
calulations of optical charge-transfer (CT) transitions at transition-metal cations \cite{DBa96},
which compare favorably with experimental data. The CT states
may be considered as excited states of trapped holes.

\section{Results and Discussion}
\label{res}
In the following discussion, in each case we start with the ionic HF
model and add on correlations afterwards to demonstrate important implications.

First, we discuss single holes trapped at TM cations. Within HF theory,
on-acceptor holes are favorable for divalent TM cations including Cu$^{2+}$,
which possess filled electron levels above the VB edge.
At trivalent Ti-site TM acceptors, on the other hand,
localized (i.e. symmetry-breaking) off-acceptor holes are highly preferred (Table 
\ref{tfeholehf}).
\begin{table}[htb]
\caption{\label{tfeholehf} \baselineskip12pt
Total HF and MP2 embedded-cluster energies for different hole states
including lattice relaxations. Energies are
renormalized to localized off-acceptor states.
$\rm O^-_O(\pi,\sigma)$: localized
off-acceptor hole in oxygen 2$p$ $\pi$- or $\sigma$-type orbitals.
$\rm (O(xy))^-$: off-acceptor hole
delocalized over the four oxygen ligands in the xy-plane. $\rm 2S+1$ denotes the
total spin multiplicity of the cluster. Note that $\rm Fe^{3+}$ and $\rm Cr^{3+}$
favor high spin ($\rm 2S_{Fe}+1 = 6$, $\rm 2S_{Cr}+1 = 4$), and $\rm Rh^{4+}$
prefers low spin ($\rm 2S_{Rh}+1 = 2$).}
\vspace{0.2cm}
\begin{tabular}{dddd}
\multicolumn{1}{c}{Defect}&\multicolumn{1}{c}{$\rm 2S+1$}&
\multicolumn{1}{c}{$\rm \Delta E^{UHF}_{total}(eV)$}&
\multicolumn{1}{c}{$\rm \Delta E^{MP2}_{total}(eV)$}\\
\hline
\multicolumn{1}{c}{$\rm Fe^{3+}_{Ti}-O^-_O(\pi)$}&\multicolumn{1}{c}{7}&
0.0&0.0\\
\multicolumn{1}{c}{$\rm Fe^{3+}_{Ti}-O^-_O(\pi)$}&\multicolumn{1}{c}{5}&0.0&
\multicolumn{1}{c}{--}\\
\multicolumn{1}{c}{$\rm Fe^{3+}_{Ti}-O^-_O(\sigma)$}&\multicolumn{1}{c}{7}&+1.1&
\multicolumn{1}{c}{--}\\
\multicolumn{1}{c}{$\rm Fe^{3+}_{Ti}-O^-_O(\sigma)$}&\multicolumn{1}{c}{5}&+1.0&
\multicolumn{1}{c}{--}\\
\multicolumn{1}{c}{$\rm Fe^{3+}_{Ti}-(O(xy))^-$}&\multicolumn{1}{c}{7}&+2.4&
+0.5\\
\multicolumn{1}{c}{$\rm Fe^{4+}_{Ti}$}&\multicolumn{1}{c}{5}&+2.4&+0.1\\
\hline
\multicolumn{1}{c}{$\rm Rh^{3+}_{Ti}-O^-_O(\pi)$}&\multicolumn{1}{c}{2}&
0.0&0.0\\
\multicolumn{1}{c}{$\rm Rh^{3+}_{Ti}-(O(xy))^-$}&\multicolumn{1}{c}{2}&+2.7&
+0.1\\
\multicolumn{1}{c}{$\rm Rh^{4+}_{Ti}$}&\multicolumn{1}{c}{2}&+0.4&$-$2.4\\
\hline
\multicolumn{1}{c}{$\rm Cr^{3+}_{Ti}-O^-_O(\pi)$}&\multicolumn{1}{c}{5}&
0.0&0.0\\
\multicolumn{1}{c}{$\rm Cr^{4+}_{Ti}$}&\multicolumn{1}{c}{2}&+1.15&$-$2.6\\
\end{tabular}
\end{table}
Here, we observe the hole localization in
$\pi$-type oxygen $2p$-orbitals (perpendicular to the oxygen-acceptor
axis). This symmetry-broken state is stabilized by defect-induced
lattice distortions and, in this way, it can acquire physical significance.
The relaxation-induced energy gain is typically 1 eV,
which emphasizes the importance of such terms. The occurrence of symmetry-broken
solutions may be considered as electronic polaron self-trapping.
Generally, hole localization
in $\sigma$-type orbitals (parallel to the oxygen-acceptor
axis) is less favorable, but deviations may occur due to the
actual electronic acceptor structure.

Interestingly, both delocalized off-acceptor holes and the formation of tetravalent
TM ions are unfavorable within HF. For example,
delocalized off-acceptor holes are by $2-3$ eV discriminated against
their localized counterparts. Similar results have been obtained in previous
HF simulations on trapped single holes
in MgO \cite{SKK86,SHG92} and NiO \cite{MJV90}. Also these calculations
predict localized off-acceptor holes to be most favorable.
In fact such holes are likely the most favorable off-acceptor solutions
in highly ionic oxides like MgO, but this behavior may prevail
in any HF calculation, thus simulating artificially ionic situations.

In {\BT} the effects of electron correlation are most significant for trivalent
TM acceptors due to the increased covalent charge
transfer from oxygen onto the TM ions. This has
important consequences: The on-acceptor hole localization is enhanced,
and, simultaneously, possible off-acceptor
holes increasingly delocalize. In contrast, at most divalent TM cations
including Cu$^{2+}$ on-acceptor holes remain stable. This behavior differs from
$\rm La_2CuO_4$ where, confirmed by our preliminary ECC, doped holes localize at the oxygen
sites. The different {\it effective} copper charges in both oxides can explain
such deviations. Further work on $\rm La_2CuO_4$ is in progress and will be published in the near
future.

Our MP2 calculations (Table \ref{tfeholehf}) illustrate that
electron correlation tends to favor
on-acceptor holes at trivalent TM acceptors.
Rh$^{4+}$ and Cr$^{4+}$ become preferred over
$\rm M-O^-$ centers, but Fe$^{4+}$ remains 0.1 eV less favorable than
$\rm Fe^{3+}-O^-$. Within DFT even Fe$^{4+}$ is stable,
and for all trivalent TM acceptors
the most favorable off-acceptor holes are delocalized at this level.

Considering off-acceptor holes,
correlation mostly affects the delocalized hole states, but localized states to
a much lesser extent. These off-acceptor-hole results
demonstrate the pronounced interplay between orbital
relaxations and electron correlation, i.e. the localized hole
states with significant HF orbital relaxations receive less correlation-energy
gain than delocalized hole states.
However, different from HF theory the energy separations between localized
and delocalized off-acceptor holes remain within some tenths eV.
This result indicates that a modest increase of ionicity could
favor localized over delocalized off-acceptor holes.
The lanthanide contraction of $5d$ TM cations might accomplish such changes
leading to observable localized
off-acceptor holes, if these are sufficiently stable against
the formation of on-acceptor holes. ESR data suggest that localized
off-acceptor holes bound to ESR-silent Pt-acceptors exist in {\BT}
\cite{VSK96}. The charge state of platinum could not be
assessed experimentally, but $\rm Pt^{4+}$ with a completely filled
5$d{\rm (\tg)}$ subshell is a possible candidate.

In summary, electron correlation essentially stabilizes the high-valent
charge states of various TM cations in {\BT}. This mechanism
resembles the Haldane-Anderson scenario \cite{HA76} in
semiconductors. In highly ionic materials, in contrast,
the charge-state stabilization has been ascribed to lattice relaxations \cite{SS81}.
The present discussion confirms that {\BT} is semi-ionic.

In comparison to most TM acceptors the hole localization properties can differ
at ionic acceptor cations due to the lack of filled acceptor levels above the VB
edge and of covalent charge transfer. Table \ref{talhole} exemplifies the
situation for $\rm Al^{3+}$. 
\begin{table}[htb]
\caption{\label{talhole} \baselineskip12pt
Hole formations at $\rm Al^{3+}_{Ti}$. Total embedded-cluster energies
are renormalized to the localized off-acceptor $\pi$-type hole state.
The results include lattice relaxation.}
\vspace{0.2cm}
\begin{tabular}{dddd}
\multicolumn{1}{c}{Defect}&
\multicolumn{1}{c}{$\rm \Delta E^{UHF}_{total}(eV)$}&
\multicolumn{1}{c}{$\rm \Delta E^{MP2}_{total}(eV)$}&
\multicolumn{1}{c}{$\rm \Delta E^{BLYP}_{total}(eV)$}\\
\hline
\multicolumn{1}{c}{$\rm Al^{3+}_{Ti}-O^-_O(\pi)$}&0.0&
0.0&0.0\\
\multicolumn{1}{c}{$\rm Al^{3+}_{Ti}-(deloc.(\pi))^-$}&+2.5&+0.4&$-$0.8\\
\multicolumn{1}{c}{$\rm Al^{3+}_{Ti}-(deloc.(\sigma))^-$}&+3.7&+1.9&
\multicolumn{1}{c}{--}\\
\multicolumn{1}{c}{$\rm Al^{3+}_{Ti}-(V_k(\pi))$}&+1.0&0.0&$-$0.4\\
\multicolumn{1}{c}{$\rm Al^{3+}_{Ti}-(V_k(\sigma))$}&+2.8&+1.6&
\multicolumn{1}{c}{--}\\
\end{tabular}
\end{table}
Obviously, on-acceptor hole
states leading to $\rm Al^{4+}$ are highly unfavorable.
ESR investigations furnished evidence in favor of $\rm V_k$ hole centers
(i.e. one hole delocalized over two neighboring ligands)
\cite{PJS92}, but there are no such indications for other acceptors in \BT.

\begin{figure}[htb]
\begin{center}
\begin{minipage}{8cm}
\epsfxsize7cm
\epsffile{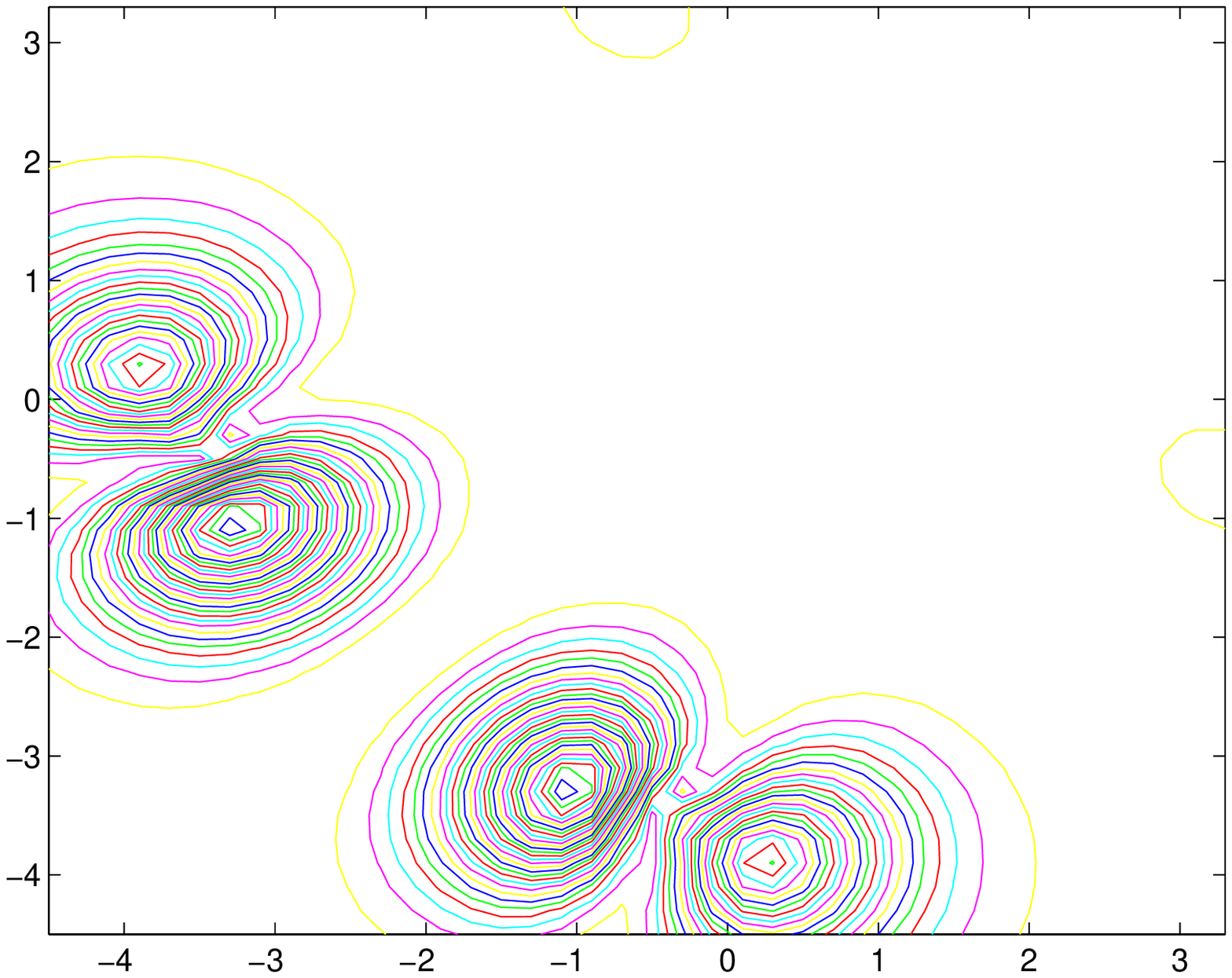}
\end{minipage}
\hspace{0.8cm}
\begin{minipage}{8cm}
\epsfxsize7cm
\epsffile{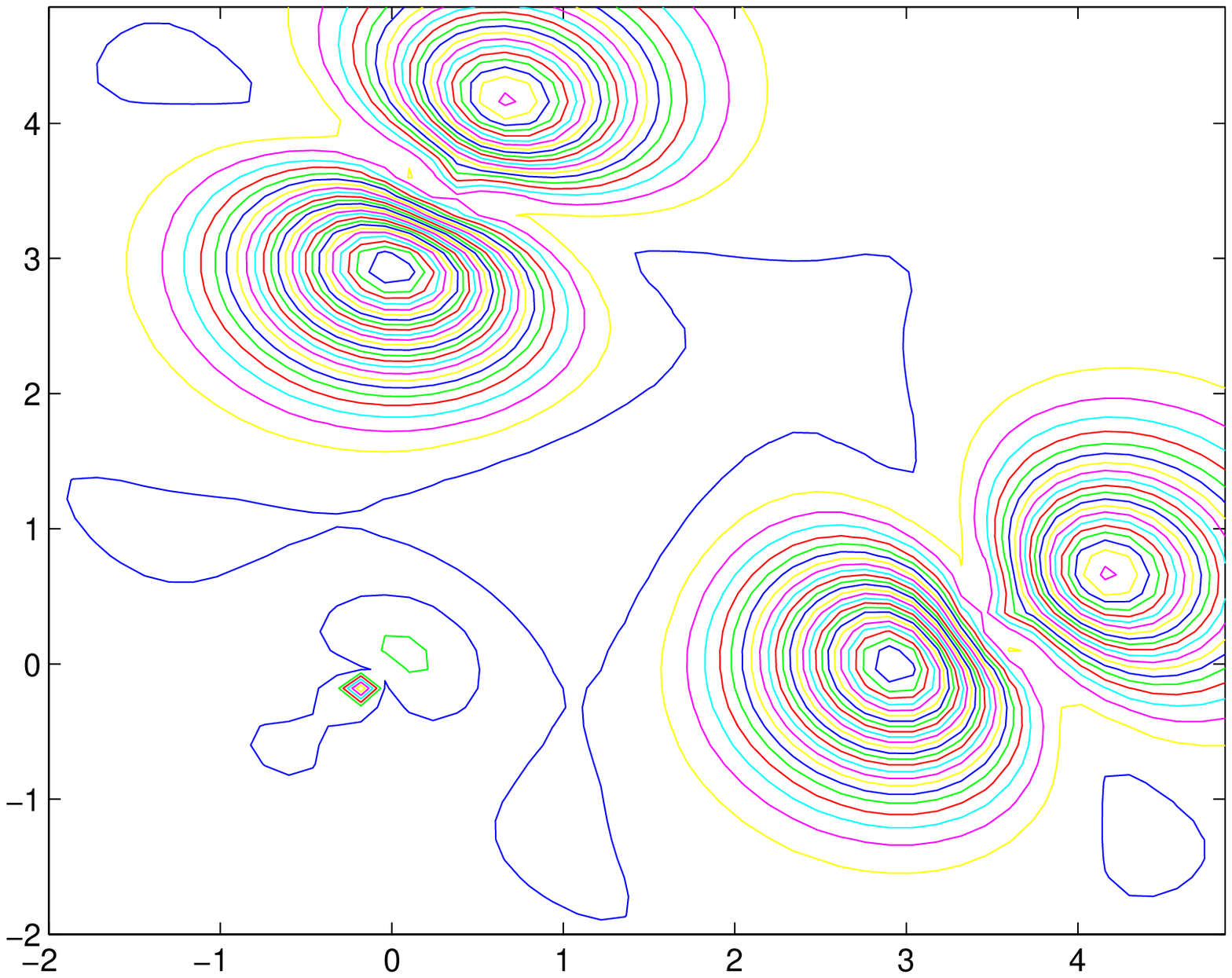}
\end{minipage}
\vspace{0.3cm}
\caption{\label{alvk} \baselineskip12pt
Hole spin density plot of $\pi$- and $\sigma$-type $\rm V_k$ centers trapped at $\rm Al^{3+}_{Ti}$.
The occupation of oxygen 2$p$ orbitals can be clearly seen. The acceptor is at (x,y)=(0,0) in
both cases. All xy-coordinates in the figures are given in atomic units. The interatomic
separation between the oxygen partners forming the $\rm V_k$ defect is 2.3 {\AA} and 2.6 {\AA}
for localization in $\pi$- and $\sigma$-type oxygen orbitals, respectively. The perfect-lattice
separation is 2.8 \AA.}
\end{center}
\end{figure}

As expected,  HF favors localized off-acceptor
$\pi$-type holes. Next favored is the $\rm V_k$ $\pi$-type hole distribution
(Fig.\ref{alvk}), but it is by 1 eV less favorable. 
The situation is even worse for completely delocalized states. Again
electron correlation supports hole delocalization.
The correlated calculations reported in Table \ref{talhole} confirm that
all hole states are of comparable energy. Most favorable are the delocalized state
and the formation of $\rm V_k$ centers. However, there is no clear
indication in favor of the latter species. But the results suggest
a delicate balance between the local electronic structure
and defect-induced lattice distortions: Increasing the acceptor ionicity would
lead to localized off-acceptor holes, whereas
a reduced ionicity implies hole delocalization. In an intermediate
small 'window' the formation of $\rm V_k$ centers might be most favorable.
This also explains why $\rm V_k$ centers are rarely observed in oxides.
Experimentally the Al-cation fits into this window, but also
the present embedded-cluster model is close to this
situation. Interestingly, $\rm V_k$ centers have been also proposed in
$\rm Al_2O_3$ \cite{KSKJ94}.

Due to reduced bonding $\sigma$-type $\rm V_k$ centers are by 1.6-1.8 eV less
favorable than $\pi$-type ones. All ion
displacements are smaller in the $\sigma$-type modification. We stress that
these results refer to Al$^{3+}$ acceptors, but
it is recalled that holes doped into
the Cu-O planes of high-T$_C$ oxides are assumed to populate
$\sigma$-type oxygen orbitals \cite{FNR89}, which, following our preliminary
calculations, is related to the occupation of copper $3d_{x^2-y^2}$ states and to
the presence of strong electron correlations.
Our present results receive particular importance, since $\rm V_k$ centers
(upon addition of a second hole) bear resemblance to small intersite
hole bipolarons. Thus, we may expect the same relaxation behavior
for hypothetical $\pi$-type and $\sigma$-type hole bipolarons
in high-T$_C$ oxides. Due to moderate
lattice distortions the latter species would be more appropriate
to coherent motions. Speculatively, $\sigma$-type bipolarons could
facilitate superconductivity, thereby supporting the scenario
of Alexandrov and Mott \cite{AM94}. Certainly, our speculations deserve further
detailed investigations.

We finally review trapped hole bipolarons in {\BT}; the details of the calculations are 
discussed extensively in Refs. 
\cite{DB95,Do95}.  Ti-site Mg$^{2+}$ is considered as the actual acceptor cation.
Two geometrical hole-acceptor configurations are compared: the bipolaron
with two holes on neighboring oxygen ions (i.e. literally a $\rm V_k$ center
with one additional hole) and
a linear complex $\rm O^--Mg^{2+}_{Ti}-O^-$. The presence of two holes introduces
a natural driving force towards localization, of which the linear
complex corresponds to minimizing the inter-hole Coulomb repulsion.
Both hole complexes have been studied under different conditions:
(1) HF treatment of the cluster employing a rigid and perfectly
structured crystal lattice. Only the actual O$^-$ partners are allowed to relax.
(2) HF description of the cluster including lattice relaxation.
(3) Correlated calculations (MP2 and DFT). To save computer capacities
further geometry optimizations
have been performed at the DFT level (LSDA and BLYP) only.

We first survey the perfect-lattice HF simulations.
Electronic interactions between the O$^{-}$ ions and nearby crystal ions
favor the spin triplet over the singlet state. Due to its antisymmetrical
charge distribution the triplet-state bipolaron benefits most from
this interaction. Increasing the bipolaron bond length rapidly
disturbs the bipolaron type states due to hole-state
delocalizations. But energetically most favorable is the linear complex. This
expected result reflects the disfavor of delocalized holes within HF.
The linear configuration with either spin is 1 eV more favorable than the
triplet-state bipolaron. 

Next we consider the effect of lattice deformations.
Most importantly, our ECC demonstrate that defect-induced lattice
relaxation and electronic correlations stabilize
hole bipolarons in {\BT}. Lattice relaxations increase the localization of
bipolaron states and favor the spin-singlet state. Thus,
such relaxations enable the possibility of embedded $\rm O^{2-}_2$
molecules analogous to isoelectronic $\rm F_2$
dimers. Noticeably, both species are unstable within HF
(e.g. see \cite{Do95}). The ultimate stability is established by
correlation-induced bonding terms. Binding energies of bipolarons can
be estimated as the energy difference $\rm E(BP)-E(O^--Mg-O^-)$, since,
due to shell-model estimates, the competitive linear complex is
only weakly bound in {\BT}. We obtain $-0.41$ eV (MP2), $-1.13$ eV (DFT-BLYP)
and $-2.1$ eV (DFT-LSDA). Calculated bond lengths are:
1.48 {\AA} (LSDA) and 1.55 {\AA} (BLYP).
As is frequently observed \cite{PF91}, LSDA overestimates bonding.
The accurate binding energy is enclosed by the BLYP- and MP2
values.

We believe that paired holes are of general importance in any oxide.
Possible differences will refer to the occupation of
$\pi$- or $\sigma$-type oxygen $2p$ orbitals giving different bonding strengths.
Finally, we emphasize that the singlet-triplet splitting of bipolaron states,
i.e. the 'spin gap', depends on the strength of lattice relaxations.
Whereas in perfect lattices the triplet state is preferred, the singlet
state becomes most favorable upon lattice distortions.
For $\pi$-type bipolarons in {\BT}, due to our MP2 calculations,
the spin gap becomes 1.50 eV in this case. 
We note that this value is essentially intrinsic to the hole bipolarons in {\BT},
and does not depend on the specific nature of the magnesium acceptor ion. In particular,
the bipolaron states are not contaminated by magnesium orbitals; we, therefore, expect a similar
spin gap for possibly existing isolated bipolarons in \BT.
In their bipolaron theory of high-T$_C$ superconductivity Mott and Alexandrov
suggested a spin gap of a few tens meV \cite{AM94}. Such a small
value would indicate that, compared with {\BT}, lattice relaxations
are present but less developed in high-T$_C$ oxides. This observation
is important if coherent motion of bipolarons is to be required.

\acknowledgements{
We gratefully acknowledge the financial support of this work by the Deutsche
Forschungsgemeinschaft (SFB 225). We also thank Prof.~O.~F.~Schirmer and
Prof.~M.~W\"ohlecke for many valuable discussions.}


\begin{thebibliography}{10}

\bibitem{PJK92}
E.~Possenriede, P.~Jacobs, H.~Kr\"ose, and O.~F. Schirmer.
\newblock {\em Appl.~Phys.} {\bf A55}, 73 (1992).

\bibitem{Ho89}
L.~Holtmann.
\newblock {\em phys.~stat.~sol. (a)} {\bf 113}, K 89 (1989).

\bibitem{VSK96}
T.~Varnhorst, O.~F. Schirmer, H.~Kr{\"o}se, R.~Scharfschwerdt, and Th.~W. Kool.
\newblock {\em Phys.~Rev.} {\bf B53}, 116 (1996).

\bibitem{AM94}
A.~S. Alexandrov and N.~F. Mott.
\newblock {\em Rep.~Prog.~Phys.} {\bf 57}, 1197 (1994).

\bibitem{An75}
P.~W. Anderson.
\newblock {\em Phys.~Rev.~Lett.} {\bf 34}, 953 (1975).

\bibitem{Sc85}
C.~Schlenker.
\newblock In D.~Adler, H.~Fritzsche, and S.~R. Ovshinsky, editors, {\em Physics
  of Disordered Materials: Mott Festschrift}, volume~3. Plenum Press, New York,
  1985.

\bibitem{DBa96}
H.~Donnerberg and R.~H. Bartram.
\newblock {\em J.~Phys.: Condens.~Matter} {\bf 8}, 1687 (1996).

\bibitem{HW85}
W.~R. Hay and P.~J. Wadt.
\newblock {\em J.~Chem.~Phys.} {\bf 82}, 270--310 (1985).

\bibitem{CAD}
R.~D. Amos and et~al., 1994.
\newblock Cambridge Analytic Derivatives Package (CADPAC) version 5.2
  (Cambridge).

\bibitem{VWN80}
S.~H. Vosko, L.~Wilk, and M.~Nusair.
\newblock {\em Can.~J.~Phys.} {\bf 58}, 1200 (1980).

\bibitem{Be88}
A.~D. Becke.
\newblock {\em Phys.~Rev.} {\bf A38}, 3098 (1988).

\bibitem{LYP88}
C.~Lee, W.~Yang, and R.~G. Parr.
\newblock {\em Phys.~Rev.} {\bf B37}, 785 (1988).

\bibitem{DB95}
H.~Donnerberg and A.~Birkholz.
\newblock {\em J.~Phys.: Condens.~Matter} {\bf 7}, 327 (1995).

\bibitem{Do95}
H.~Donnerberg.
\newblock {\em J.~Phys.: Condens.~Matter} {\bf 7}, L689 (1995).

\bibitem{LC86}
G.~V. Lewis and C.~R.~A. Catlow.
\newblock {\em J. Phys. Chem. Solids} {\bf 47}, 89 (1986).

\bibitem{CM82}
C.~R.~A. Catlow and W.~C. Mackrodt, editors.
\newblock {\em Computer Simulation of Solids}, volume 166 of {\em Lecture Notes
  in Physics}.
\newblock Springer Verlag, Berlin, Heidelberg, New York, 1982.

\bibitem{Le83}
M.~Leslie.
\newblock {\em Solid State Ionics} {\bf 8}, 243 (1983).

\bibitem{VHH84}
J.~M. Vail, A.~H. Harker, J.~H. Harding, and P.~Saul.
\newblock {\em J.~Phys. C: Solid State Phys.} {\bf 17}, 3401 (1984).

\bibitem{Ku78}
A.~B. Kunz and D.~L. Klein.
\newblock {\em Phys.~Rev.} {\bf B17}, 4614 (1978).

\bibitem{SKK86}
A.~L. Shluger, E.~A. Kotomin, and L.~N. Kantorovich.
\newblock {\em J.~Phys. C: Solid State Phys.} {\bf 19}, 4183 (1986).

\bibitem{SHG92}
A.~L. Shluger, E.~N. Heifets, J.~D. Gale, and C.~R.~A. Catlow.
\newblock {\em J.~Phys.: Condens.~Matter} {\bf 4}, 5711 (1992).

\bibitem{MJV90}
J.~Meng, P.~Jena, and J.~M. Vail.
\newblock {\em J.~Phys.: Condens.~Matter} {\bf 2}, 10371 (1990).

\bibitem{HA76}
F.~D.~M. Haldane and P.~W. Anderson.
\newblock {\em Phys.~Rev.} {\bf B13}, 2553 (1976).

\bibitem{SS81}
A.~M. Stoneham and M.~J.~L. Sangster.
\newblock {\em Phil.~Mag.} {\bf B43}, 609 (1981).

\bibitem{PJS92}
E.~Possenriede, P.~Jacobs, and O.~F. Schirmer.
\newblock {\em J.~Phys.: Condens.~Matter} {\bf 4}, 4719 (1992).

\bibitem{KSKJ94}
L.~Kantorovich, A.~Stashans, E.~Kotomin, and P.~W.~M. Jacobs.
\newblock {\em Int.~J.~Quant.~Chem.} {\bf 52}, 1177 (1994).

\bibitem{FNR89}
J.~Fink, N.~N\"ucker, H.~A. Romberg, and J.~C. Fuggle.
\newblock {\em IBM~J.~Res.~Develop.} {\bf 33}, 372 (1989).

\bibitem{PF91}
P.~Fulde.
\newblock {\em Electron Correlations in Molecules and Solids}, volume 100 of
  {\em Solid-State Sciences}.
\newblock Springer Verlag, Berlin, Heidelberg, 1991.

\end{thebibliography}
\end{document}